\documentclass[nonacm, sigconf]{acmart}
\renewcommand\footnotetextcopyrightpermission[1]{}

\usepackage{url}
\usepackage{breakurl}
\usepackage{booktabs} 
\usepackage{balance}
\usepackage{graphicx}
\usepackage{listings}
\usepackage{xspace}
\usepackage{color}
\usepackage{tabularx}
\usepackage{amsfonts}
\usepackage{amsmath}
\usepackage{cleveref}
\usepackage{syntax}
\usepackage{enumitem}

\usepackage{subfig}
\usepackage[export]{adjustbox}
\usepackage[skip=2ex]{caption}

\usepackage[]{hyperref}

\definecolor{forestgreen}{RGB}{14,189,14}

\newcommand{\todo}[1]{}
\newcommand{\ds}[1]{}
\newcommand{\gagan}[1]{}
\newcommand{\Rishikesh}[1]{}
\newcommand{\shachee}[1]{}
\renewcommand{\todo}[1]{{\color{blue}\textbf{TODO:} #1}}
\renewcommand{\ds}[1]{{\color{teal}\textbf{[ds]:} #1}}
\renewcommand{\gagan}[1]{{\color{violet}\textbf{[gagan]:} #1}}
\renewcommand{\shachee}[1]{{\color{orange}\textbf{[shachee]:} #1}}
\renewcommand{\Rishikesh}[1]{{\color{forestgreen}\textbf{[rishikesh]:} #1}}

\newcommand{\enterprise}{Microsoft\xspace}

\usepackage[utf8]{inputenc}
\usepackage[scaled=0.75]{beramono}

\lstdefinelanguage{Lang}{}
\lstdefinelanguage{YAML}{}
\begin{document}
\title{Continuous Benchmark Generation for Evaluating Enterprise-scale LLM Agents}

\author[Saxena, et. al.]{Divyanshu Saxena$^{1}$, Rishikesh Maurya$^{2}$, Xiaoxuan Ou$^{2}$, Gagan Somashekar$^{2}$, Shachee Mishra Gupta$^{2}$\newline
Arun Iyer$^{2}$, Yu Kang$^{2}$, Chetan Bansal$^{2}$, Aditya Akella$^{1}$, Saravan Rajmohan$^{2}$ \newline
{\normalsize $^{1}$The University of Texas at Austin, $^{2}$Microsoft}
}

\begin{abstract}
The rapid adoption of AI agents across  domains has made systematic evaluation crucial for ensuring their usefulness and successful production deployment.
Evaluation of AI agents typically involves using a fixed set of benchmarks and computing multiple evaluation metrics for the agent.
While sufficient for simple coding tasks, these benchmarks fall short for enterprise-scale agents, where services and requirements evolve continuously and ground-truth examples are sparse.
We propose a process of benchmark generation that helps evolve the benchmarks as the requirements change and perform robust evaluation of evolving AI agents.
We instantiate this approach for a case study of service migration from one deployment platform to another at a large public enterprise.
Our approach relies on semi-structured documents where developers express the high-level intent, and uses state-of-the-art LLMs to generate benchmarks from just a small number of such documents.
Overall, this process results in a maintainable evaluation framework, enabling rapid feedback on agent performance and facilitating targeted improvements.
\end{abstract}

\maketitle

\section{Introduction}

Modern enterprises increasingly rely on large language model (LLM)–based agents to automate software engineering and operational workflows.
Examples include \emph{build and CI/CD agents}~\cite{logsage-cicd-agents} that adapt build configurations and monitor CI/CD pipelines for issues, \emph{incident management and triaging agents}~\cite{comet-issre24,yu2025triangle} that route incidents to the appropriate service owners, and \emph{troubleshooting agents}~\cite{t2c-osdi25} that help reproduce failures and synthesize fixes.
These enterprise-scale agents operate across thousands of services and heterogeneous artifacts such as source code, deployment manifests, configuration files, and documentation.
Their outputs directly affect production environments—incorrect actions can impact operations, ranging from user dissatisfaction and degraded service quality to complete unavailability of critical services.

While several evaluation mechanisms and benchmarks exist for general-purpose coding agents designed for isolated and generic tasks (e.g., function synthesis or bug fixing), enterprise-scale LLM agents pose fundamentally different evaluation requirements.
They must \textit{reliably} produce \textit{high-quality} outputs, even as the task requirements change.
Moreover, these agents themselves evolve -- operators continually integrate newer model versions and reasoning capabilities—making evaluation an ongoing necessity rather than a one-time exercise.

Existing benchmarks, however, are ill-suited to this setting.
Most are static, targeting a fixed and narrow set of tasks. When requirements change, they must be manually revised, often resulting in numerous variants (e.g., SWE-bench~\cite{swe-bench} and its extensions~\cite{swe-bench-lite,swe-bench-lite-S,swe-bench-verified}).
While useful for reproducibility and enabling transparent evaluation, they fail to capture the continuous evolution typical of enterprise environments.
Their rigidity makes longitudinal evaluation difficult and imposes a recurring maintenance burden.

We identify two key challenges in evaluating enterprise-scale agents.
First, these agents operate under changing dynamics, e.g., deployment platforms, operational policies, or troubleshooting guides may change frequently.
An agent that performs well today may degrade as dependencies evolve or organizational requirements shift.
Second, the heterogeneity of context and artifacts makes defining ground truth difficult. 
Conventional benchmarks typically rely on well-defined, static reference outputs, but enterprise-scale tasks span diverse file types, programming languages, and infrastructure components -- all of which evolve independently, e.g. for feature additions and bug fixes.
Consequently, determining what constitutes the correct or intended outcome for a given task is non-trivial, often requiring significant effort to carefully extract the right ground truths and thus, imposing a recurring maintenance burden.

We propose a new approach to evaluating enterprise-scale agents.
Instead of relying on fixed, static benchmarks, we introduce a {\it continuous benchmark generation pipeline} that evolves alongside changing requirements without imposing significant developer burden on creating these benchmarks.
In this paper, we ground the discussion for one such use case from our experience at a major cloud provider \enterprise (anonymized for review), where services were migrated to a new infrastructure platform.

\begin{table*}[t!]
\small
\centering
\begin{tabular}{p{2.5cm} p{4.5cm} p{9.5cm}}
\toprule
\textbf{File Type} & \textbf{Reason} & \textbf{Example} \\
\midrule
Script files & File path formats differ across operating systems & Change directory paths from forward slash (``/'') in Linux to back slash (``\textbackslash'') in Windows \\
\addlinespace
Source and build files & Change in supported libraries & `System.Drawing.Common' is a .NET library that provides access to Windows Graphics Device Interface. SDC does not exist on Linux systems and must be updated to cross-platform libraries like ImageSharp or SkiaSharp \\
\addlinespace
Dependency configuration files & Different installation or deployment setups across OSes & Jaeger, a popular monitoring library needs to be deployed as a Windows service on Windows, but as a container on Linux/Kubernetes \\
\addlinespace
Service orchestration files & Changes in requirements from the platform management teams & Modify or add new \texttt{Dockerfile}s for successful deployment. \\
\bottomrule
\end{tabular}
\caption{Examples of file-level changes required during service migration.}
\label{tab:examples}
\end{table*}

\noindent\textbf{Motivating Use Case: Service Migration Agents.}\quad
Enterprise services are typically deployed on shared infrastructure platforms, which themselves evolve over time as underlying components are modified or replaced.
We observed one such instance over the past six months in which existing services were transitioned to new infrastructure platform.
Such platform-level changes are often intrusive, requiring the dependent services to undergo corresponding migrations.
For example, when the underlying operating system changes (e.g., transitioning between Windows and Linux Operating Systems), build and compilation scripts must be updated to reflect the new environment.
Similarly, if the service relies on OS-specific libraries, such as the System Drawing package (see example in Table~\ref{tab:examples}) that is only functional for Windows, the source code must be updated to use cross-platform libraries (e.g., ImageSharp). 
As is the case for enterprise-scale agents, this service migration task spanned several heterogeneous artifacts, including build and compilation scripts, deployment files, and even source code. We discuss examples of these changes in more detail in Section~\ref{sec:case-study}.

\noindent\textbf{Our Proposal: Continuous Benchmark Generation.}\quad
Our vision is to develop a methodology that allows benchmarks to be evolved as operational and organizational requirements change, and enables thorough evaluation of agents over a diverse set of tasks and services.
We achieve this via two key components.

The first of these are \emph{developer-authored semi-structured Knowledge Bases (KBs)}.
The term KB is generally used to denote resources used by LLM agents~\cite{stackfeed,wu2024stark,kblam}, but in our context, it describes the specifics of the evaluation task at hand, such as the goal, context, and expected outcomes for the task.
We envision that a task (e.g., service migration) can be specified as a suite of {\it KB documents} -- one document per sub-task (e.g., update the logging library in the service code).
Crucially, any updates to the requirements can be captured simply by editing an existing KB document or adding a new one. 
Moreover, by developing distinct documents for heterogeneous artifacts observed in enterprise-scale services, we can get fine-grained metrics about agent performance. 

The second component in our proposal are \emph{reference implementations} derived from a small number of real commits that serve as ``gold-standard'' examples.
By linking these examples to the corresponding KB documents, we can construct the ground truths that can be used in the benchmark.
As services evolve and new migrations are completed -- whether manually or automatically -- new commits can be incorporated to expand and refine the benchmark. 

Overall, our framework can enable longitudinal, adaptive evaluation of enterprise-scale agents as both the agents and the underlying platforms evolve.
In the remainder of this paper, we (1) present our case study of platform migration at \enterprise (2) describe our vision for continuous benchmark generation~(\Cref{sec:benchmark-generation}); and (3) discuss broader implications of using our approach~(\Cref{sec:future-directions}).

\section{Case Study: Service Migration}
\label{sec:case-study}

The deployment environment we worked with hosts thousands of services serving over hundreds of millions of users.
Performance and efficiency are critical for hosting and developing services at this scale.
In an effort to achieve higher performance standards and good cluster efficiency, infrastructure teams continuously upgrade various pieces of the platform.
Further, services at large enterprises may need to migrate from one platform to another, in search of better performance and efficiency trade-offs, resulting from improved hardware capabilities and up-to-date software stack support.

\noindent{\bf Service Migration Challenges.}\quad
Modifications to the infrastructure components may involve changes transparent to the services, e.g., changing the infrastructure to add a new generation of memory devices, but they may often also involve changes that are more intrusive in nature.
For instance, one such platform migration at \enterprise required switching the underlying Operating System (e.g., between Windows OS and Linux).
This, in turn, required updates to various build and source code files.
We provide a representative list of such changes and examples in \Cref{tab:examples}.

These changes are often needed across a heterogeneous set of artifacts because of several reasons.
First, there are some fundamental differences between the two Operating Systems, requiring changes to how file paths are formatted, which dependencies are used, and how services are built (see \Cref{tab:examples}).
Secondly, platform management teams may also specify requirements on which libraries or dependencies they can support, and various best practices to optimally manage services.
Thirdly, these changes often result in modifications to deployment files as well, because any changes in the dependencies must be reflected in the associated deployment files, such as Dockerfiles.

 

\Cref{tab:examples} highlights a representative subset of the several classes of changes needed for this task.
In the specific platform migration at \enterprise, such changes were needed across a diverse set of services.
Given the scale of service migration needed for our use case, we sought LLM agents for migration, aiming to automate repetitive tasks and accelerate planning.
To put into perspective the extent of the complexity, a small set of 11 services that had been migrated manually by developers, took nearly 20-25 weeks per service on average and a total of over 350 PRs. 

\begin{figure*}[!t]
    \centering
    \includegraphics[width=0.75\linewidth]{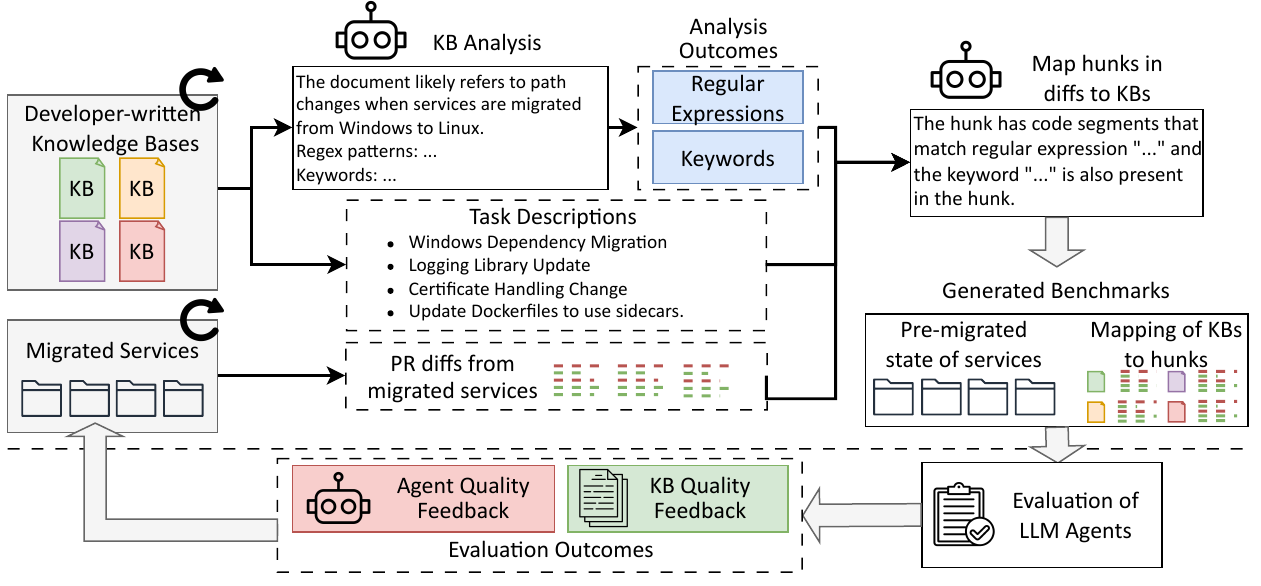}
    \caption{A pipeline to generate benchmarks from manually migrated repos.}
    \label{fig:pipeline-architecture}
\end{figure*}

\noindent{\bf Benchmarking Needs.}\quad
There are a wide variety of language models and tools that one can use to develop agents.
Given the heterogeneous nature of the artifacts requiring migration, benchmarking is necessary to understand how well state-of-the-art agents perform, and which change patterns (such as the ones shown in \ref{tab:examples}) are best suited for which agents.

To evaluate, we construct an initial evaluation dataset based on the 11 manually migrated services: for each of those, the benchmark consists of the pre-migrated state of the service repo (that will be passed to the agent), and the oracle patch (diff between the pre- and post-migrated states of the repo).

However, just this evaluation set was insufficient for three main reasons.
First, the evaluation set reflected idiosyncratic changes, specific to the services rather than systematic ones that can be generalized across services.
This is because at the enterprise-level, services still undergo feature updates, bug fixes, etc. {\it while} platform migration is happening.
For instance, a deployment file update required the addition of environment variables to specify the underlying Operating System and the respective compilation toolchain to use, however the commit from a service included additional environment variables that were later found to be service-specific.
Second, benchmarks based on fixed set of services fail to capture the breadth of deployment heterogeneity because not all 11 services used all parts of the new platform.
For example, some services adopted containerization with Docker, thus using Dockerfile and Kubernetes manifests for deployment, while others continued using legacy PowerShell (.ps1) and shell (.sh) scripts for deployment. 
This resulted in a mix of containerized and non-containerized services within the same migration wave.
Finally, simply evaluating each agent and model on the entire dataset made it hard to extract the performance sliced by the change patterns -- the same commit may include changes for more than one change patterns, even the same file may have changes across multiple change patterns! 
This motivated the design of a new benchmark generation methodology tailored for enterprise-scale, continuous migration tasks.
\section{Continuous Benchmark Generation Pipelines}
\label{sec:benchmark-generation}

To address the limitations of existing static and manually-labor intensive benchmarks, our proposal is to develop \textit{continuously-evolving benchmark generation pipelines} that automatically produces benchmarks, from migrated services given a set of high-level task description documents.
The key principle is a clean separation between {\em what to test} (requirements specification) and {\em how to test} (i.e., produce concrete instances for evaluation).

In the remainder of this section, we describe our vision -- a pipeline that uses natural language documents and migration-related commits.
Figure~\ref{fig:pipeline-architecture} depicts the overall architecture.

\noindent\textbf{High-level Overview.}\quad
At a high level, the pipeline takes the tuple $\mathcal{I}=(\mathcal{S}, \mathcal{T})$ as inputs, where $\mathcal{S}$ represents a small set of migrated services that act as the foundation for constructing the benchmark, and $\mathcal{T}$ represents a set of task description documents prepared by developers (more details in Section~\ref{subsec:kbs}).
The output is the set $\mathcal{O}=\{(\triangleleft_{s}, t, \mathcal{H}_{s, t}) \forall s \in \mathcal{S}, t \in \mathcal{T}\}$, where $\triangleleft_{s}$ denotes the pre-migration state of the service $s$, and $\mathcal{H}_{s, t}$ denotes diff hunks from migration-related commits of service $s$, relevant for the task $t$.

\subsection{Knowledge Bases}\label{subsec:kbs}
At the heart of our proposal are developer-authored documents (denoted by the set $\mathcal{T}$ in the above formulation) describing the various tasks that must be performed by the AI agent to achieve its objectives.
We call these task description documents Knowledge Bases (KBs) that can be thought of as external knowledge sources that can be used by LLM agents~\cite{stackfeed}.

Building these Knowledge Bases requires developers to break the high-level objective into smaller tasks, each describing a specific and related set of changes.
For instance, for the high-level objective of service migration that we studied in Section~\ref{sec:case-study}, sub-tasks could be `update the logging dependencies based on the Operating System' or `update the Kubernetes manifest files to utilize the latest recommended sidecar frameworks'. 
In our approach, a Knowledge Base is a set of documents, where each document maps to a narrow sub-task like the ones described above.

Capturing requirements in the form of KBs imposes little developer burden -- these requirements are described even today in the form of migration manuals or troubleshooting guides.
These KBs can then evolve naturally as platforms deprecate features, introduce new APIs, or modify resource specifications.
This makes them a durable anchor point for continuous benchmark generation, as we discuss below.
This intent-based system design has been a common insight in several other domains as well~\cite{intent-based-design-pacmi25}.

\subsection{Benchmarks from KBs and Migrated Repos}
The key challenge in using migrated enterprise-level services as benchmarks is in isolating migration-related changes from idiosyncratic service updates and ensuring that outdated changes do not compromise benchmark validity.
Our insight is that all the needed changes for migration, are captured in the set of task descriptions $\mathcal{T}$, and hence, only those changes from the migrated services must be used in the benchmark that map to some task in the set $\mathcal{T}$.

To perform this mapping, we rely on two techniques:
First, if the KB document explicitly provides keywords (for instance, in a separate section in the document) that must be present in a relevant code segment, we can directly search for these keywords and map each hunk in a diff that has those keywords to this document. These keywords are used by the agents to identify "where" the changes are to be made. 
Secondly, KB documents may not have precise keywords, but rather descriptions of what `lines of code' to look for.
For instance, for the file path format migration (see Table~\ref{tab:examples}), the KB may specify `Windows drive names' implying that we must look for strings matching drive name patterns (such as `C:\textbackslash' or `D:\textbackslash').
In this case, we can leverage the reasoning and text-generation capabilities of LLMs to generate regular expressions for these types of descriptions.

\noindent\textbf{Evolving the Benchmarks.}\quad
Any updates to the requirements, can be simply incorporated into the KB documents.
Further, this process of benchmark generation can also act as feedback for KB quality itself -- say a new edit was made to the KB, but the benchmarks do not show the update despite having a commit for it, then one can infer that the KB content needs to be updated.
Similarly, the benchmark can be expanded as more services are migrated, and more migration-related commits are available.

\subsection{Evaluation Metrics}

Our benchmarks are constructed of code diffs for each provided task.
We can use this benchmark to evaluate agents by computing the following metrics:

{\em 1. Line-edit Precision and Recall.}
We can compare the agent’s line edits with the ground-truth edits produced by our benchmark pipeline. Each predicted edit is matched to a ground-truth edit within a fixed edit distance.
Line-edit precision measures the fraction of correctly predicted edits, while line-edit recall measures the fraction of required edits successfully performed by the agent.

{\em 2. Per-KB Precision and Recall.}
While line-edit metrics capture syntactic fidelity, they do not reveal which tasks the agent handled well or poorly. To complement them, we evaluate correctness at the KB granularity. An agent is said to cover a KB if at least one of its predicted changes matches the ground truth for that KB. Per-KB precision measures the fraction of covered KBs that were actually relevant to the service, and per-KB recall measures the fraction of required KBs that the agent successfully covered.

We believe more metrics can provide richer feedback to the pipeline, and is an interesting direction of research. 


\begin{table}[t]
\small
\centering
\begin{tabular}{p{1.5cm} c c c}
\toprule
\textbf{Repository} & \textbf{Precision} & \textbf{Recall} & \textbf{F1 Score} \\
\midrule
Repo1 & 1 & 0.5 & 0.66 \\
Repo2 & 1 & 0.4 & 0.57 \\
Repo3 & 1 & 0.667 & 0.8 \\
Repo4 & 1 & 0.25 & 0.4 \\
\bottomrule
\end{tabular}
\caption{Benchmark Generation Results}
\label{tab:results}
\end{table}

\subsection{Evaluating the Benchmarks}

We implemented the proposed pipeline and evaluated it using 137 KB documents across 4 services.
For comparison, a group of developers manually constructed benchmarks based on migration-related commits from the same services.
We then assessed how well the automatically generated benchmarks aligned with these manually designed ones. As shown in \Cref{tab:results}, the pipeline-generated benchmarks achieved high precision and strong recall.

Interestingly, the automated benchmarks also proved to be more reliable.
For instance, when evaluated against the first version of the manually-designed benchmarks, we saw low recall scores.
A deeper inspection revealed that the manually written benchmarks included obsolete files that were still present in the repositories.
Since our pipeline derives ground truth directly from reference commits, such files are automatically excluded, yielding cleaner and more accurate benchmarks.
\section{Discussion and Future Directions}
\label{sec:future-directions}

In this paper, we presented a case study for the evaluation of LLM agents at enterprise-scale.
Using the example of platform migration induced service migration, we showcase why evaluation of enterprise-scale agents is inherently complex and continuous.
Our proposal is a pipeline that can generate benchmarks, as requirements evolve or more services are available as ground-truth.

Our approach addresses two key challenges. First, it leverages long-term commit histories to build realistic benchmarks without manually filtering feature or bug-fix changes. Second, it minimizes developer effort by reusing existing natural-language migration documents to automatically construct benchmarks.

Our proposal opens several directions for future work. A key question is how to assess the accuracy of generated benchmarks, requiring new mechanisms and metrics for correctness evaluation.
Another is how to use benchmarking feedback to better prompt agents -- for instance, one can provide targeted feedback if the LLM agent performs poorly on specific tasks.
At enterprise scale, one could even deploy multiple agents specialized for different task types.
We believe this evaluation methodology may extend beyond service-migration agents studied in this paper -- for instance, to troubleshooting agents leveraging natural-language guides -- though this remains a direction for future exploration.

\label{EndOfPaper}
\balance
\bibliographystyle{ACM-Reference-Format}
\bibliography{refs}

\end{document}